\documentclass[12pt,a4paper,dvips]{article} \usepackage{a4p}
\usepackage{cite,mcite} \usepackage{graphicx} \usepackage{physics }
\usepackage{epsfig,rotating,lscape } \usepackage{l3_title,ifthen} \usepackage{Lep}
\Lep{2}
\journalname{Phys. Lett. B}
\date{August 2, 2002}
\preprint{2002-064}
\newlength{\capwidth}
\newcommand{\EE}{\mathrm{e}^+\mathrm{e}^-}
\newcommand{\MM}{\mu^+\mu^-}
\newcommand{\TT}{\tau^+\tau^-}
\newcommand{\LL}{\ell^+\ell^-}
\newcommand{\FF}{\mathrm{f}\bar{\mathrm{f}}^\prime}

\newcommand{\QQ}{\mathrm{q}\bar{\mathrm{q}}^\prime}
\newcommand{\qq}{\mathrm{q}\bar{\mathrm{q}}}

\newcommand{\WW}{\mathrm{W}^+\mathrm{W}^-}
\newcommand{\Wen}{\mathrm{e^+} \nu_\mathrm{e} \mathrm{W^-}}
\newcommand{\Wln}{\mathrm{W^-} \to \mathrm{\ell^-} \bar{\nu}_\ell}
\newcommand{\Wqq}{\mathrm{W^-} \to \mathrm{q}\bar{\mathrm{q}}^\prime}
\newcommand{\epl}{\mathrm{e^+}}
\newcommand{\emi}{\mathrm{e^-}}
\newcommand{\nel}{\nu_\mathrm{e}}

\newcommand{\ane}{\bar{\nu}_\mathrm{e}}
\newcommand{\anm}{\bar{\nu}_\mu}
\newcommand{\ant}{\bar{\nu}_\tau}
\newcommand{\anl}{\bar{\nu}_\ell}
\newcommand{\enu}{\mathrm{e^+} \nu_\mathrm{e}}
\newcommand{\ean}{\mathrm{e^-} \bar{\nu}_\mathrm{e}}
\newcommand{\lan}{\mathrm{\ell^-} \bar{\nu}_\mathrm{\ell}}

\newcommand{\pz}{\phantom{0}}
\newcommand{\pzz}{\phantom{00}}

\newcommand{\afig}{{\kern -0.25em a}}
\newcommand{\bfig}{{\kern -0.25em b}}
\newcommand{\cfig}{{\kern -0.25em c}}
\newcommand{\dfig}{{\kern -0.25em d}}

\begin{document}
\begin{titlepage}
  
  \title{Production of Single W Bosons at LEP
    and \\
    Measurement of \boldmath$\rm W W \gamma$ Gauge Coupling
    Parameters}
  
  \author{L3 Collaboration}
  
  \begin{abstract}
    
    Single W boson production in electron-positron collisions is
    studied with the L3 detector at centre-of-mass energies between
    $192\GeV$ and $209\GeV$.  Events with two acoplanar hadronic jets
    or a single energetic lepton are selected, and the single W cross
    section is measured. Combining the results with measurements at
    lower centre-of-mass energies, the ratio of the measured cross
    section to the Standard Model expectation is found to be
    $1.12^{+0.11}_{-0.10}\pm0.03$. From all single W data, the
    WW$\gamma$ gauge coupling parameter $\kappa_\gamma$ is measured to
    be $1.116^{+0.082}_{-0.086}\pm0.068$.
  \end{abstract}

  \submitted

\end{titlepage}

\section {Introduction}

At LEP, single W production\footnote{The charge conjugate reactions
  are understood to be included throughout this Letter.}, $\EE \to
\Wen$, provides one of the best experimental measurements of the
trilinear gauge boson coupling parameters, in particular of the
coupling parameter $\kappa_\gamma$~\cite{tsukamoto}.  It is
  complementary to the measurement of the gauge boson coupling
  parameters in W pair production. In the single W
process, only the electromagnetic couplings of the W boson are probed,
unlike in W pair production which is also sensitive to the couplings
between W and Z bosons. The single W cross section depends only on the
$\kappa_\gamma$ and $\lambda_\gamma$ parameters~\cite{tsukamoto} which
are related to the magnetic dipole moment, $\mu_\mathrm{W} =
(e/(2\,\MW)) \left(1 + \kappa_\gamma + \lambda_\gamma \right)$, and
the electric quadrupole moment, $\mathrm{q}_\mathrm{W} = (-e/\MW^2)
\left(\kappa_\gamma - \lambda_\gamma \right)$, of the W boson.  An
accurate measurement of these couplings constitutes a crucial test of
the Standard Model of electroweak interactions~\cite{gsw,veltman},
that has been made in previous studies by the LEP
experiments~\cite{l3wen1,l3wen2,alephwen1,alephwen2,delphiwen,delphitgc}.

The Standard Model predictions are, at tree level,
$\kappa_\gamma = 1$ and $\lambda_\gamma = 0$. Higher order contributions
are small~\cite{tgcloop} compared to the measurement precision at LEP.
Deviations from the Standard Model prediction would thus indicate
anomalous corrections or an internal structure of the W boson.

A particular feature of single W production is a final state positron
scattered at very low polar angle, which remains undetected.  Thus the
detector signature of this process is two hadronic jets and a large
transverse momentum imbalance, in case of hadronic W
decays, or a single energetic lepton for leptonic W decays.

In this Letter the measurements of the cross sections of single W
boson production at centre-of-mass energies $\sqrt{s}=192-209\GeV$ are
presented.  Combining the results with those obtained at lower
centre-of-mass energies~\cite{l3wen2}, the ratio of the measured cross
section to the Standard Model expectation is determined and
$\kappa_\gamma$ and $\lambda_\gamma$ are measured.

\section {Data and Monte Carlo Samples}

The data were collected with the L3 detector~\cite{l3det} at LEP at
several mean centre-of-mass energies as detailed in
Table~\ref{tab:list}.  They correspond to an integrated luminosity of
$452.6\,\pb$. The separate luminosities at the six energy points are
also given in Table~\ref{tab:list}.

For signal studies, samples of $\EE \to \enu \FF$ events are generated
using both the GRC4F~\cite{grc4f} and the EXCALIBUR~\cite{excalibur}
Monte Carlo generators.  For background studies the following Monte
Carlo programs are used: KORALW~\cite{koralw} ($\EE \to \WW \to {\rm
  f\bar{f}^\prime f^{\prime\prime}\bar{f}^{\prime\prime\prime}}$),
KK2F~\cite{kk} and PYTHIA~\cite{pythia} ($\EE \to \qq(\gamma)$),
KK2F ($\EE \to \MM(\gamma), \; \TT(\gamma)$),
KORALZ~\cite{koralz} ($\EE \to \nu \bar{\nu}(\gamma)$),
BHAGENE3~\cite{bhagene} and BHWIDE~\cite{bhwide} for large angle
Bhabha scattering ($\EE \to \EE(\gamma)$), TEEGG~\cite{tee} for small
angle Bhabha scattering ($\EE \to \EE \gamma$), DIAG36~\cite{diag36}
and PHOJET~\cite{phojet} for leptonic and hadronic two-photon
processes, respectively, and GRC4F and EXCALIBUR for other 4-fermion
final states not listed above.

The response of the L3 detector is simulated with the GEANT
program~\cite{geant}, which takes into account the effects of energy
loss, multiple scattering and showering in the detector. The GHEISHA
program~\cite{gheisha} is used to simulate hadronic interactions in
the detector. Time dependent detector inefficiencies are taken into
account in the simulation.

\section {Signal Definition}

The single W signal is defined from $\EE \to \enu \FF$ Monte Carlo
events that satisfy the following phase-space
requirements~\cite{l3wen1,l3wen2}: 

\begin{eqnarray}
  | \, \rm{\cos\theta_{e^+}} |        & > & 0.997             \nonumber   \\
  \label{eq:eqn02}
  \mbox{min}(E_\mathrm{f},E_{\bar{\mathrm{f}}^\prime}) & > & 15   \GeV  \\ 
  | \, \rm{\cos\theta_{e^-}} |        & < & 0.75  \;\;\;\;\; \rm{for} \;
\enu \ean \; \rm{events \; only,}                             \nonumber
\end{eqnarray}
where $\rm{\theta_{e^+}}$ is the polar angle of the outgoing positron,
and $E_\mathrm{f}$ and $E_{\bar{\mathrm{f}}^\prime}$ are the fermion
energies. Generated $\rm \EE \to \enu \FF$ events that do not satisfy
these conditions are considered as background. They come mostly from
the reaction $\EE \to \WW$.  Inside the phase-space
region~(\ref{eq:eqn02}), 82\% of the events have an invariant mass of
the $\FF$ pair, $m_{\FF}$, such that
$|\mathrm{m}_{\FF}-m_\mathrm{W}|<3\,\Gamma_\mathrm{W}$, where
$m_\mathrm{W}$ and $\Gamma_\mathrm{W}$ are the mass and the width of
the W boson~\cite{mw-pdg}, thus indicating a high signal purity.

Signal cross sections are calculated, within the above phase-space
definition, using the Monte Carlo generators GRC4F and EXCALIBUR. The
latter is also used to determine selection efficiencies for the signal
process and to reweight Monte Carlo events for the extraction of the
gauge couplings.  The main difference between the two generators is in
the treatment of the masses of fermions, which are taken to be
massless in EXCALIBUR.  The theoretical uncertainty on the predictions
for the single W production cross section is estimated to be
5\%~\cite{lep4f}.  This includes the effect of using a smaller
electromagnetic coupling to account for the low momentum transfer of
the photon in single W production and taking into account QED
radiative corrections expected for a $t$-channel process.

\section {Analysis}

Events with two hadronic jets and large transverse momentum imbalance
and events with single energetic electrons, muons or taus are
selected. The selection criteria are optimised for different
centre-of-mass energies separately.  In the following, the analyses at
energies above $\sqrt{s}=202\GeV$ are described in detail.

\subsection {Hadronic Final States}

Candidates for the hadronic decay of single W bosons are identified as
high multiplicity hadronic events containing two acoplanar jets and no
isolated leptons. The energy deposition in the electromagnetic
calorimeter must be greater than $15\GeV$ and the total visible energy
must be in the range: $0.30 < E_{vis}/\sqrt{s} < 0.65$. The transverse
energy of the event is required to be greater than $0.2\, E_{vis}$.
These criteria efficiently remove fermion-pair and hadronic two-photon
background.

All energy clusters in an event are combined into two hadronic jets
using the DURHAM jet clustering algorithm~\cite{durham}. To further
reject events from the radiative process $\EE \to {\rm q \bar{q}}
(\gamma)$, the angle between the missing momentum vector and the beam
axis is restricted to $|\cos\theta_{miss}| < 0.92$.  In
addition, the acoplanarity between the two jets must be larger than
$11^\circ$.

In order to suppress background from the $\EE \to \WW$ process where
one of the W bosons decays into leptons, events containing electrons,
muons or photons with high energy are rejected.

Three jets are formed for every remaining event.  The solid angle,
$\Omega$, defined by the directions of these jets is required to be
less than 4.8~srad. This criterion removes part of the remaining
$\tau^+\nu_\tau \, \QQ$ final states with the $\tau$ lepton decaying
hadronically.  Events with $\tau$-jets are further removed by
constructing a probability to identify the best candidate for a narrow
$\tau$-jet, based on cluster and track multiplicity, as well as on the
mass and the momentum of the jet.

\begin{figure} [ht]
  \begin{center}
    \mbox{\epsfig{file=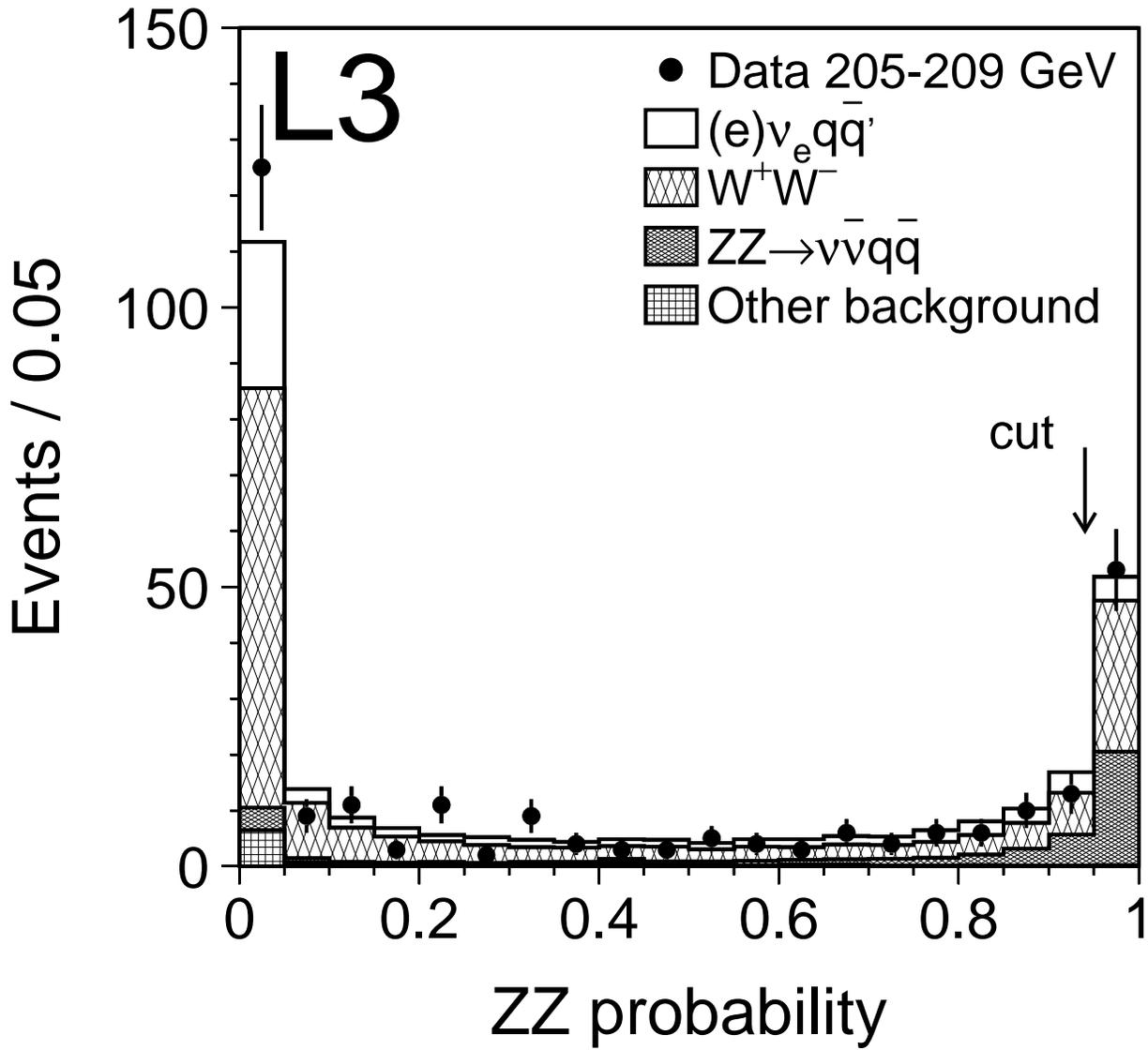,width=1.0\textwidth}}
  \end{center}
  \caption{\label{fig:zz}
    Distribution of the ZZ probability for the selected hadronic
    events above  $\sqrt{s}=202\GeV$ and Monte Carlo
    expectations.  The arrow indicates the position of the applied
    cut.}
\end{figure}

Z boson pair production in which one Z boson decays hadronically and
the other into a pair of neutrinos can mimic the signal signature.  A
ZZ probability is constructed using the following quantities: the
velocity, the invariant mass and the opening angle of the dijet
system, the missing momentum, and the reconstructed neutrino energy
assuming single W kinematics.  A cut on this probability, shown in
Figure~\ref{fig:zz}, efficiently removes this background.

The numbers of events selected at each centre-of-mass energy are listed
in Table~\ref{tab:list}, together with the selection efficiencies and
the Standard Model expectations, calculated with EXCALIBUR.

\begin{figure} [ht]
  \begin{center}
    \mbox{\epsfig{file=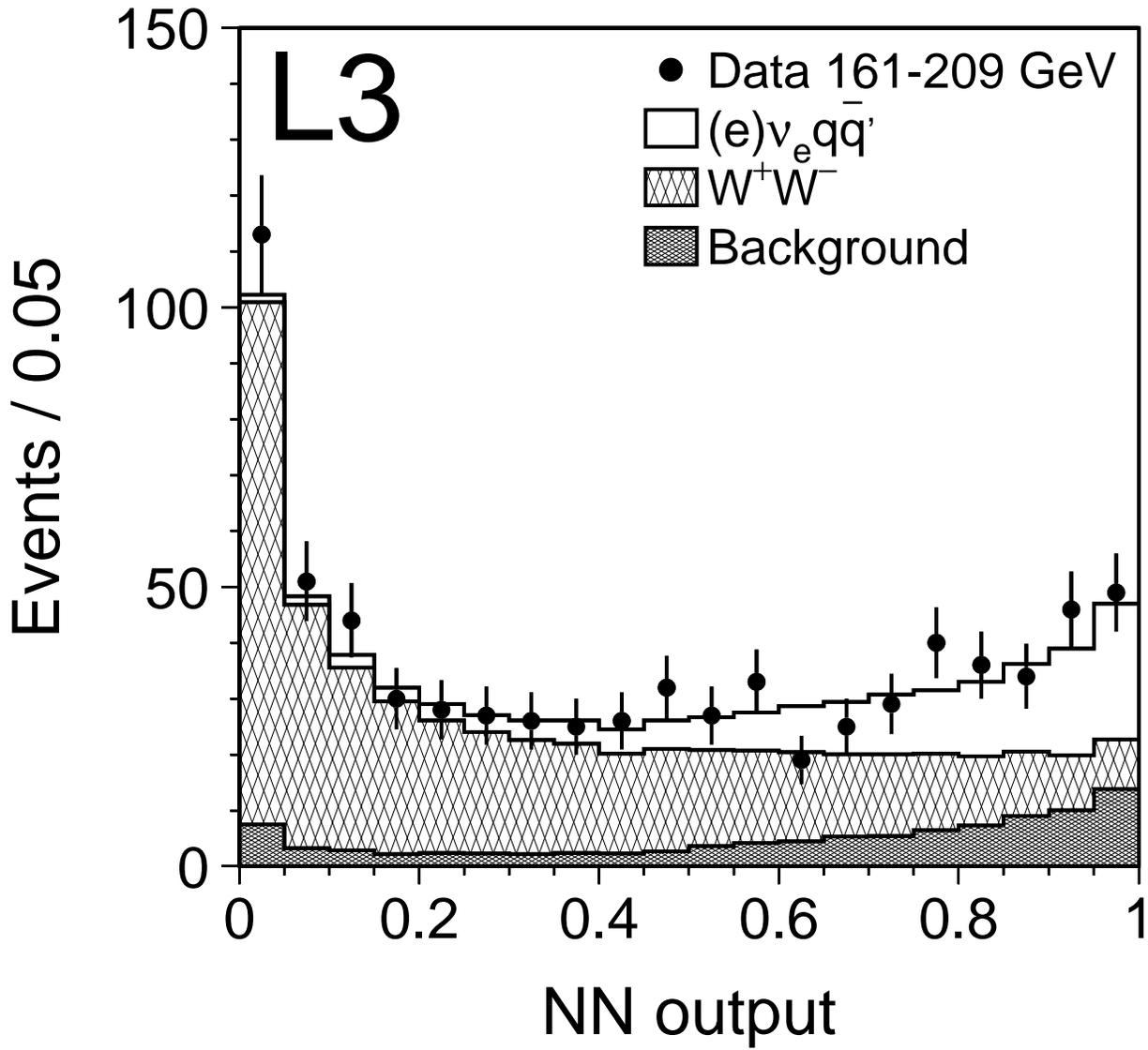,width=1.0\textwidth}}
  \end{center}
  \caption{\label{fig:nn}
    Distribution of the output of the neural network, used to identify
    hadronic single W decays. The data collected at
    $\sqrt{s}=161-209\GeV$ are shown, together with the background
    contributions and the expected signal.}
\end{figure}

In order to further differentiate between the signal and the $\EE \to
\WW$ background, a discriminating variable is constructed using a
neural network approach~\cite{nnet}. The inputs to the neural network
include three classes of variables.  Global quantities are used, such
as the velocity of the detected hadronic system, calculated as the
ratio of the missing momentum and the visible energy, and the visible
invariant mass. Variables based on a 2-jet topology are included, like
the sum of the masses of the two jets, the ratio of the mass and the
energy of the most energetic jet, the reconstructed energy of the
neutrino, assuming single W kinematics, the missing momentum, the
rescaled invariant mass and velocity of the hadronic system, and the
angle between the two jets. Finally, variables assuming a 3-jet
topology are considered: the solid angle $\Omega$, the DURHAM
parameter $y_{23}$ for which the number of jets in the event changes
from two to three, and the minimal opening angle between any two jets.
Figure~\ref{fig:nn} shows the output of the neural network used in the
subsequent analysis.

\subsection {Leptonic Final States}

Single W candidates where the W boson decays leptonically have the
distinct signature of one high energy lepton and no other significant
activity in the detector. Events with one charged lepton identified
either as electron, muon or hadronic $\tau$-jet~\cite{l3wen2} are
selected. Events containing well measured tracks that are not associated to
the lepton are rejected.

Several selection criteria are applied to suppress background from
two-fermion production $\EE\to\LL(\gamma)$. The angle between the
lepton candidate and any track or calorimetric object that could be
assigned to a second particle in the opposite hemisphere is required
to be less than 2.8 rad for electron and muon candidates and less than
2.4 rad for hadronic tau candidates. Furthermore, the visible mass of
all energy clusters must be less than $0.1\sqrt{s}$. No more than
$10\GeV$ are allowed to be deposited in the low angle calorimeters.

In single electron final states, the electron energy must exceed 92\%
of the total energy, calculated as the sum of the lepton energy and
the energies of all neutral clusters in the event. The polar angle is
restricted to the central detector region,
$|\cos{\theta_\mathrm{e}}|<0.75$. These requirements reduce the
contribution from Bhabha and Compton scattering and from the process
$\EE \to \EE \nu \bar{\nu}$ where the $\EE$ pair originates from a
low-mass virtual photon. Converted photons from the process $\EE \to
\nu \bar{\nu} \gamma$ might fake a single electron.  Since
configurations with the $\nu \bar{\nu}$ pair originating from a Z
boson are preferred, the mass recoiling against the single electron
candidate is required to be incompatible with the Z boson mass and to
exceed $0.48\sqrt{s}$.

For single muon final states, the muon energy, measured in the muon
chambers and in the central tracker, is required to be greater than
90\% of the total energy.  The fiducial volume for this analysis is
defined to be $|\cos\theta_\mu| < 0.86$. Additional requirements are
put on the missing transverse momentum, $p^{miss}_\bot\ge
0.08\sqrt{s}$, and on the mass recoiling against the muon,
$M_{rec}/\sqrt{s}\le 0.91$.

Single tau candidates are accepted in a polar angular range of
$|\cos\theta_\tau|<0.75$.  The number of charged tracks reconstructed
in the central tracking system and associated with the hadronic tau
must be either 1 or 3.  Background is further reduced by requiring the
mass recoiling against the tau to be in the range: $0.55\le
M_{rec}/\sqrt{s}\le 0.93$.

\begin{figure} [ht]
  \begin{center}
    \mbox{\epsfig{file=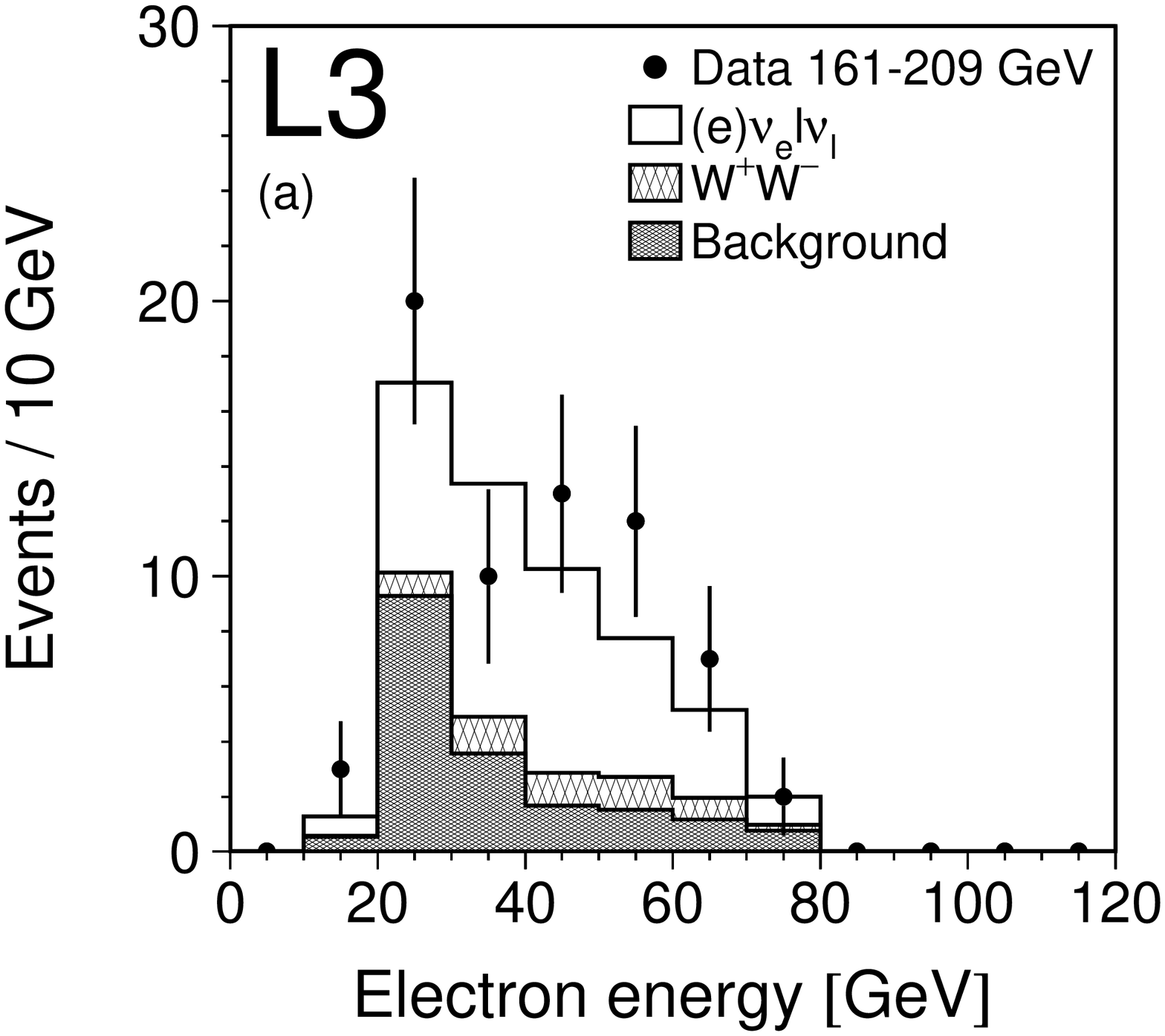,width=0.49\textwidth}}
    \mbox{\epsfig{file=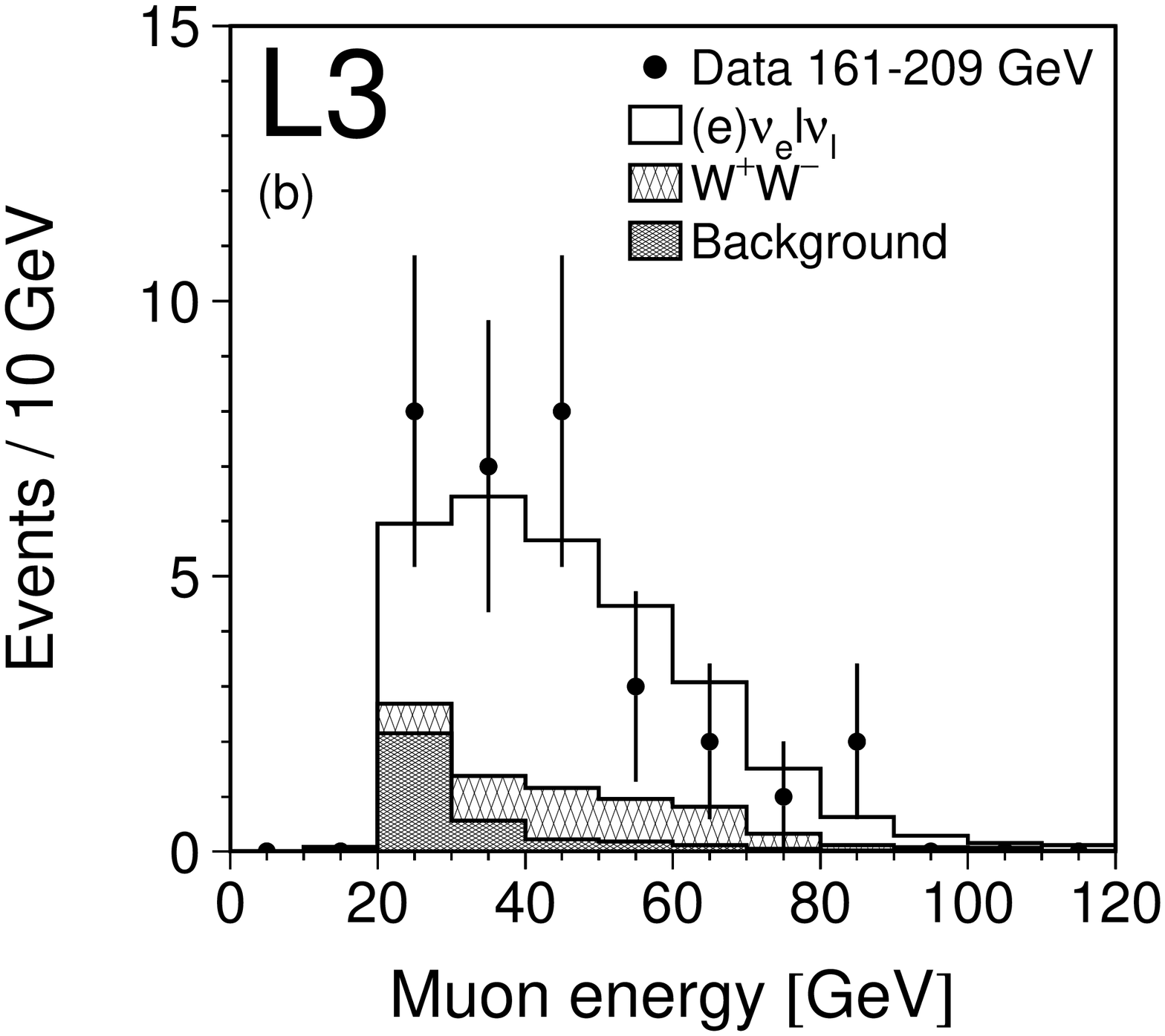,width=0.49\textwidth}}
    \mbox{\epsfig{file=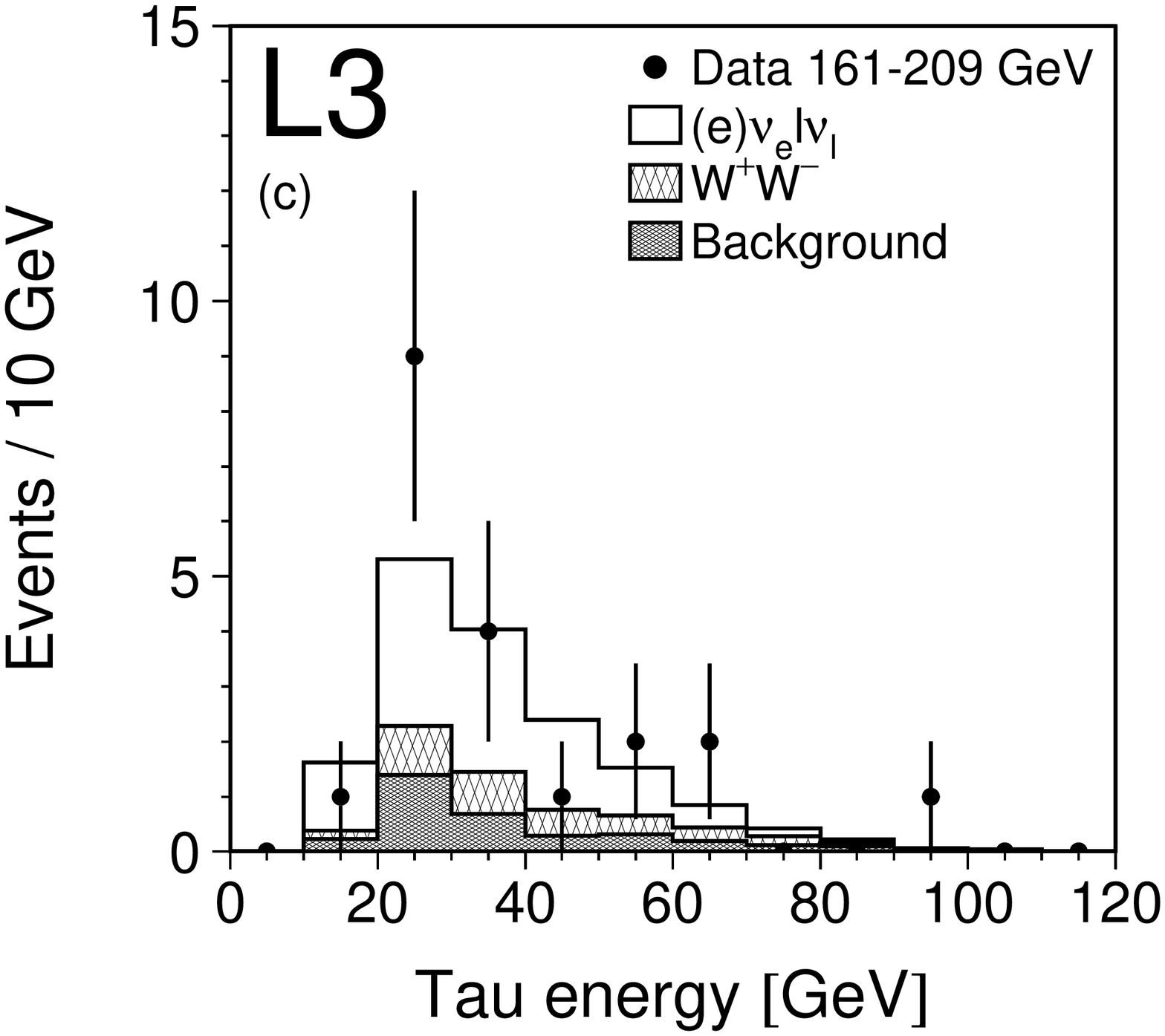,width=0.49\textwidth}}
    \mbox{\epsfig{file=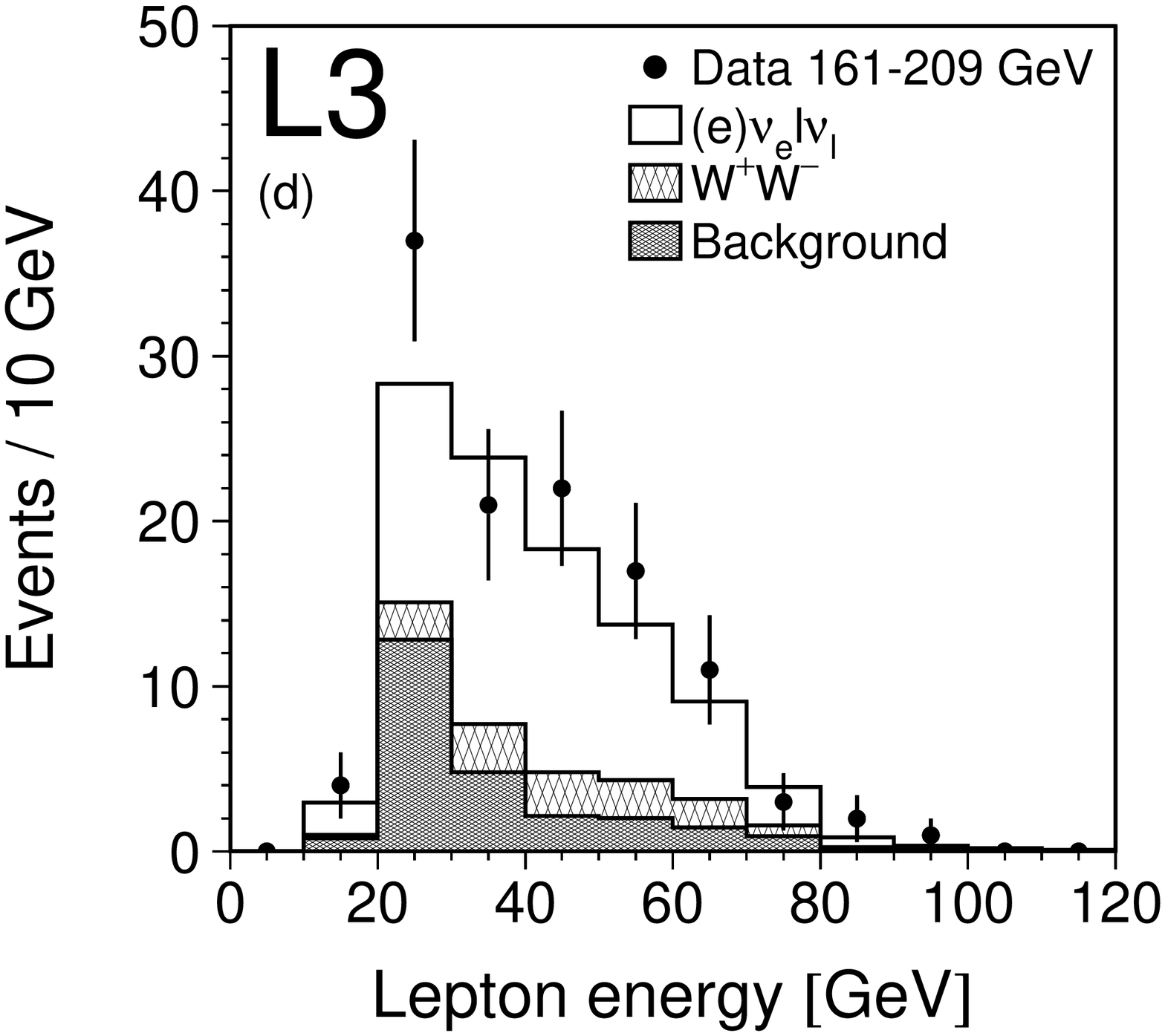,width=0.49\textwidth}}
  \end{center}
  \caption{\label{fig:lp}
    The energy spectrum of the lepton candidates, selected as
    (a) electrons, (b) muons or (c) hadronic $\tau$-jets, and their
    sum (d). Data
    measured at $\sqrt{s}=161-209\GeV$ are presented, together with
    Monte Carlo expectations.}
\end{figure}

The trigger efficiencies are determined directly from data in a sample of $\EE
\to \WW \to \ell^+ \nu_\ell \, \ell^{\prime\,-} \bar{\nu}_{\ell^\prime} $ events to be
$(93\pm3)\%$, $(88\pm2)\%$, and $(97\pm3)\%$ for the electron, muon and tau channels,
respectively. The numbers of observed and expected events as well as
the selection efficiencies are summarised in Table~\ref{tab:list}.
Figure~\ref{fig:lp} shows the lepton energy spectra for the selected
events.

\section {Cross Section Measurement}

The cross section of the signal process at each energy point is
determined by a binned maximum likelihood fit to the distributions of
the neural network output in the hadronic decay channel and of the
combined lepton energy distributions in the lepton channel. The
background shapes and normalisations are fixed to the Monte Carlo
prediction.

\begin{figure} [ht]
  \begin{center}
    \mbox{\epsfig{file=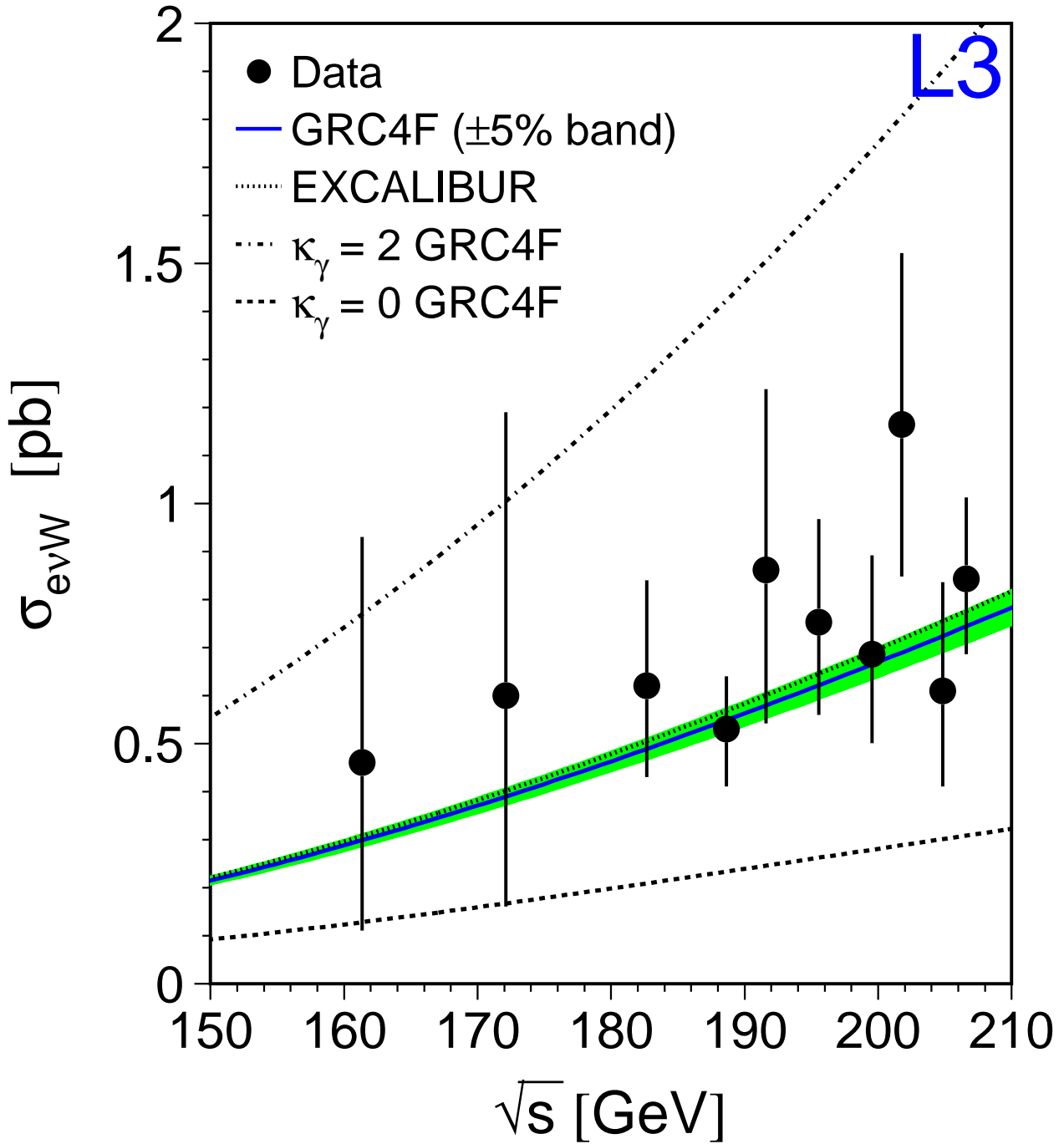,width=1.0\textwidth}}
  \end{center}
  \caption{\label{fig:xsband}
    The measured cross section of single W production as a function of
    $\sqrt{s}$. 
    The solid and dotted lines show predictions of the GRC4F and
    EXCALIBUR Monte Carlo programs, using the Standard Model value of
    $\kappa_\gamma=1$.  A $\pm 5\%$ band illustrates the theoretical
    uncertainty~[25]. Possible deviations from the Standard Model for
    $\kappa_\gamma=0$ and $\kappa_\gamma=2$ are shown by the dashed
    and dash-dotted curves. }
\end{figure}

The measured signal cross sections for the phase space region
(\ref{eq:eqn02}) are summarised in Table~\ref{tab:xsection} for the
six centre-of-mass energies. When combining the hadronic and leptonic
channels, Standard Model values for the branching fractions of the W
boson~\cite{wbr} are assumed. The measured cross section values are
consistent with the Standard Model expectations calculated with GRC4F
and EXCALIBUR.  The dependence of the cross section on the
centre-of-mass energy agrees well with the predictions, as shown in
Figure~\ref{fig:xsband}.

The systematic uncertainties on the cross section measurements for the
hadronic and leptonic channels are summarised in
Table~\ref{tab:systematics}. A significant contribution arises from
the difference between the GRC4F and EXCALIBUR signal modelling,
estimated by comparing the signal efficiencies obtained with the two
Monte Carlo programs.

In the hadronic channel the uncertainty due to the choice of the
neural network structure is tested by changing the parameters of the
network. Effects of detector resolution and calibration are studied by
smearing and shifting the kinematic variables that are fed into the
network. They give a negligible contribution to the systematic
uncertainty. The identification of leptons is studied using control
data samples of two-fermion production and differences between data
and the simulation are taken into account in the systematics. For
leptons, the uncertainties on the trigger efficiencies are included.

Limited Monte Carlo statistics introduce uncertainties on the signal
efficiency and the expected background levels. In addition, the $\WW$
and ZZ background cross sections are varied within the uncertainties
on the theoretical predictions of 0.5\% and 2\%~\cite{lep4f},
respectively.  As a cross-check, a fit of the $\WW$ cross section is
performed, keeping the single W contribution fixed to the Standard
Model prediction. It agrees, within the statistical accuracy, with the
expectation for $\WW$ production.  Finally, a variation of the bin
sizes of the fitted distributions is taken into account.

The results at different centre-of-mass energies are further analysed
in terms of the ratio, $R$, of the measured cross section,
$\sigma^{meas}_{\e \nu \W}$, to the theoretical expectation,
$\sigma_{\e \nu \W}^{theo}$, calculated with GRC4F. The $R$ value is
extracted by combining the individual likelihood functions of the
cross section measurements.  Systematic uncertainties and correlations
between them are taken into account in the combination. Uncertainties
on the background cross sections are treated as correlated between all
data sets.  Systematics originating from the signal modelling are
taken as correlated between energy points, but uncorrelated between
the hadronic and leptonic channels. Also the uncertainties on the
trigger efficiencies for leptons are treated as correlated between
energy points. All other systematic contributions are assumed to be
uncorrelated.

A fit to all data at $\sqrt{s}=161-209\GeV$ yields
\begin{eqnarray*}
  R & = & \sigma^{meas}_{\e \nu \W}/\sigma_{\e \nu
  \W}^{theo} =
  1.12^{+0.11}_{-0.10}\pm0.03 \, ,
\end{eqnarray*}
where the first uncertainty is statistical and the second systematic.
Good agreement of the cross section measurements with the Standard
Model expectation is found.

\section {\boldmath$\rm W W \gamma$ Gauge Couplings}

Figure 4 shows the sensitivity of the single W cross section to
anomalous values of $\kappa_\gamma$.  A binned maximum likelihood fit
to the neural network output distributions and the lepton energy
spectra is used to extract $\kappa_\gamma$ and $\lambda_\gamma$.  In
the fit, each Monte Carlo event is assigned a weight that depends on
the generated event kinematics and the values of $\kappa_\gamma$ and
$\lambda_\gamma$.  The dependence of the W pair background on the
gauge couplings is also taken into account.

Assuming custodial $SU(2) \times U(1)$ gauge symmetry, the Z boson
gauge couplings $g_1^\mathrm{Z}$, $\kappa_\mathrm{Z}$ and
$\lambda_\mathrm{Z}$ are constrained to: $\kappa_\mathrm{Z} =
g_{1}^{\mathrm{Z}} - \tan^2 \theta_{\rm w} \times (\kappa_\gamma - 1)$
and $\lambda_\mathrm{Z} = \lambda_\gamma$.  In addition, the weak
charge of the W bosons is assumed to be one, $g_{1}^{\mathrm{Z}} = 1$.
These constraints are applied in the fit, but affect only the
background contributions, as the signal process depends on $\rm
\kappa_\gamma$ and $\rm \lambda_\gamma$ only.

Similar systematic error sources as for the cross section
determination are studied for the coupling measurement. The dominant
systematic uncertainty arises from the difference in the signal
efficiency estimated using the GRC4F and EXCALIBUR Monte Carlo
generators. The effect on $\kappa_\gamma$ and $\lambda_\gamma$ is
found to be 0.047 and 0.063, respectively. Both programs agree on the
ratio of cross sections with and without anomalous values of the gauge
couplings.

The theoretical uncertainty of 5\%~\cite{lep4f} on the total cross section for
single W boson production translates into a systematic variation of
0.042 for $\kappa_\gamma$ and 0.010 for $\lambda_\gamma$.  The
influence of the uncertainties~\cite{lep4f} on the $\WW$ and ZZ cross
section predictions is found to be 0.002 on $\kappa_\gamma$ and 0.010
on $\lambda_\gamma$. 

The systematic uncertainties due to the signal modelling and the
background estimation are taken as correlated between the different
data sets. Systematic effects arising from limited Monte Carlo
statistics, event selection and detector description are assumed to be
uncorrelated between the individual channels and centre-of-mass
energies. These effects mainly affect the overall normalisation of the
cross sections in the individual data sets.

\begin{figure} [ht]
  \begin{center}
    \mbox{\epsfig{file=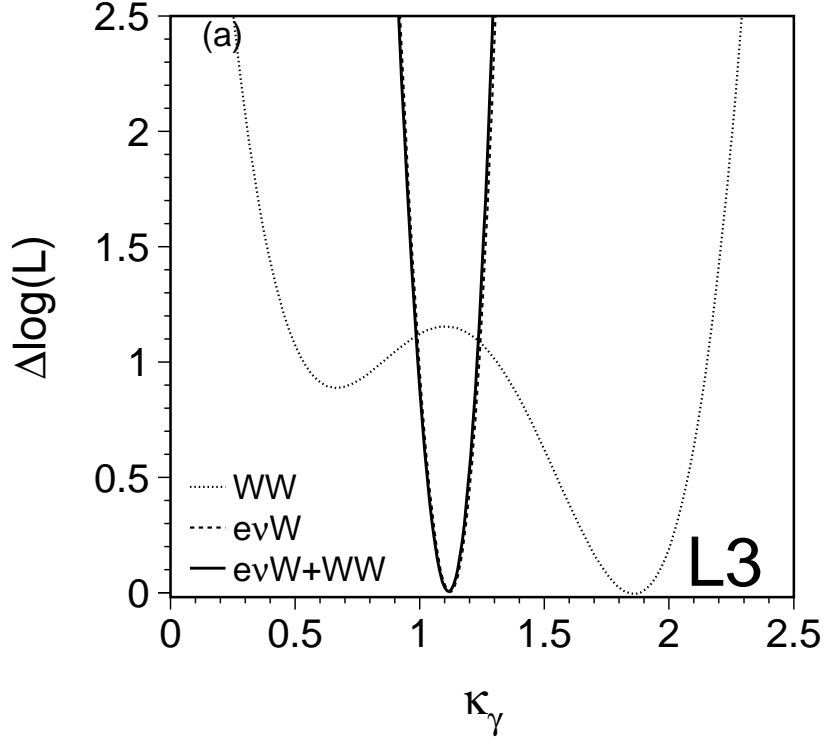, width=0.65\textwidth}}\\
    \mbox{\epsfig{file=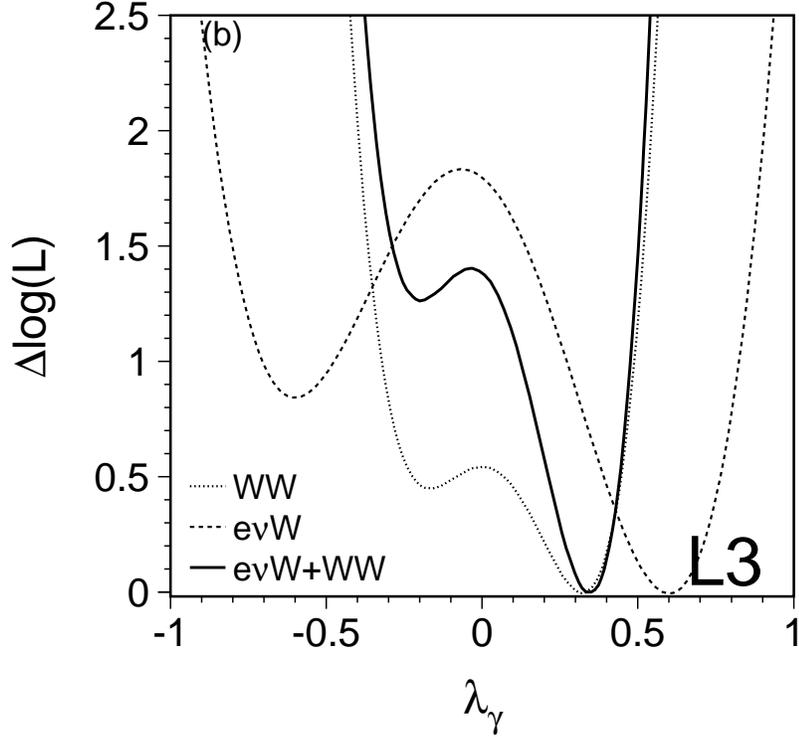,width=0.65\textwidth}}
  \end{center}
  \vspace{-10mm}
  \caption{\label{fig:fits}
    Dependence of the negative log-likelihood function, $\Delta
    \log(\mathrm{L})$, on the WW$\gamma$ gauge couplings (a)
    $\kappa_\gamma$ and (b) $\lambda_\gamma$. In each case the other
    coupling is fixed in the fit to its Standard Model value. For
    comparison, the likelihood functions are shown for the individual
    contributions of the signal and the $\WW$ background. Again, in
    each case the other process is fixed to its Standard Model
    expectation. Systematic uncertainties are taken into account.}
\end{figure}

Single W production is particularly sensitive to the gauge coupling
$\kappa_\gamma$. The parameter $\lambda_\gamma$ is therefore set to
zero in the fit for $\kappa_\gamma$.  Combining the new data with
those collected at $\sqrt{s}=161-189\GeV$~\cite{l3wen2}, yields:
\begin{eqnarray*}
  \kappa_\gamma  & = & 1.116^{+0.082}_{-0.086}\pm 0.068\, .
\end{eqnarray*}
This result agrees well with the Standard Model prediction of unity.
The likelihood distributions, shown in Figure~\ref{fig:fits}~\afig,
demonstrate that the single W data dominates the determination of
$\kappa_\gamma$.  The limits on $\kappa_\gamma$ at 95\% confidence
level are:
\begin{eqnarray*}
  0.90 \; < &  \kappa_\gamma & < \; 1.32 \,.
\end{eqnarray*}

\begin{figure} [ht]
  \begin{center}
    \mbox{\epsfig{file=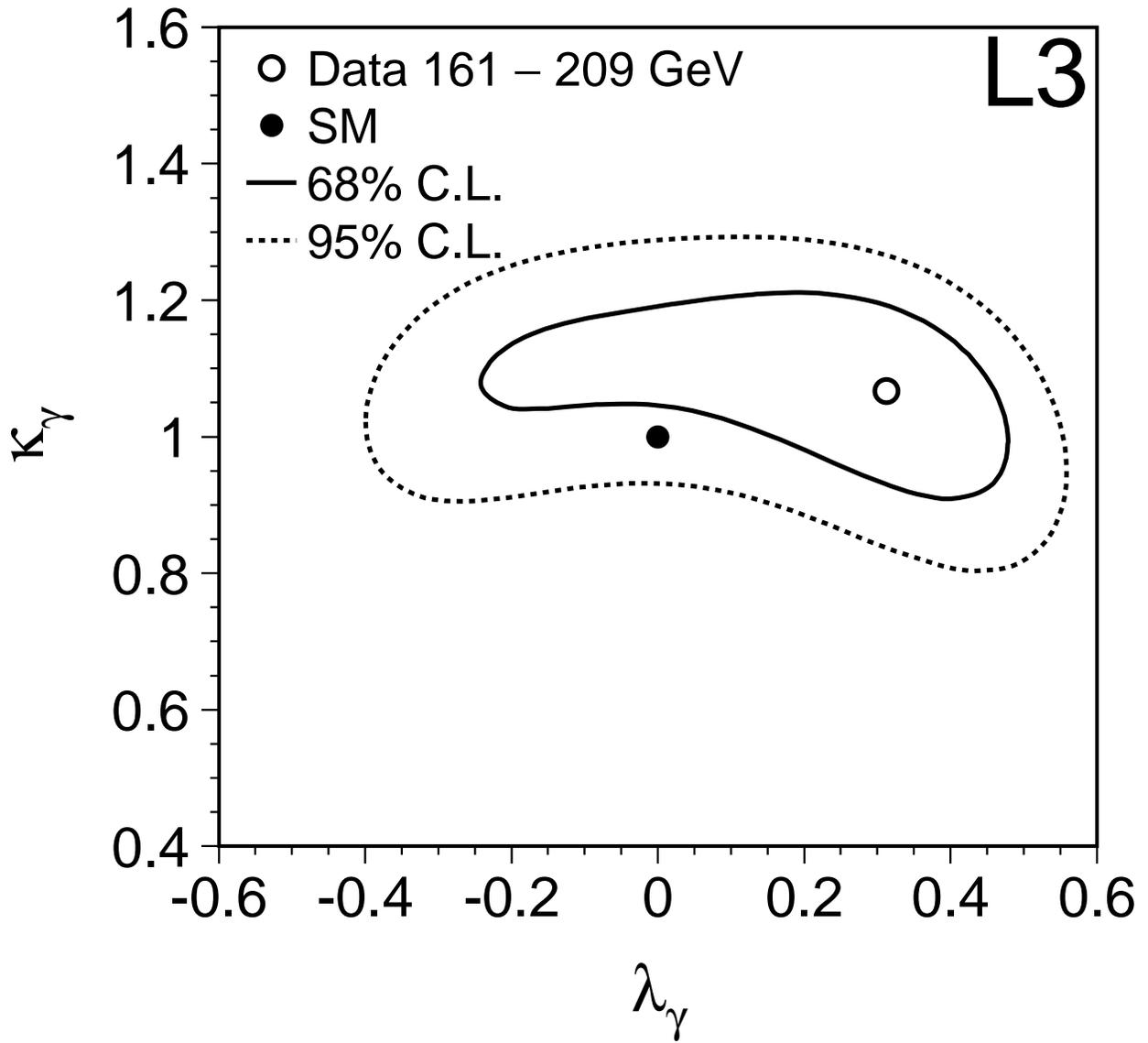,width=1.0\textwidth}}
  \end{center}
  \caption{\label{fig:cn}
    The contours corresponding to 68\% and 95\% confidence level
    regions in the $\rm \kappa_\gamma - \rm \lambda_\gamma$ plane. The
    result of the fit and the Standard Model prediction are also
    shown. Systematic uncertainties are taken into account.}
\end{figure}

Unlike the measurement of $\kappa_\gamma$, the determination of
$\lambda_\gamma$ is mainly driven by a variation of the $\WW$
background and less by the single W signal, as illustrated in the
likelihood distributions shown in Figure~\ref{fig:fits}~\bfig. 
When $\kappa_\gamma$ is fixed to the Standard Model value one, the
following results for $\lambda_\gamma$ are obtained:
\begin{center}
  \begin{tabular}{rcr@{}l@{}l@{\hspace*{2cm}}r c l}
    $\lambda_\gamma$ & = & $0.35$ & ${}^{+0.10}_{-0.13}$ & $\pm 0.08$ &
    $-0.37 \; <$ & $\lambda_\gamma$ & $< \; 0.61 \;\; (95\% \,\mathrm{C.L.}) $
    \,.
  \end{tabular}
\end{center}
Finally, varying both couplings $\kappa_\gamma$ and $\lambda_\gamma$ freely in
the fit yields:
\begin{center}
  \begin{tabular}{rcr@{}l@{}l@{\hspace*{2cm}}r c l}
    $\kappa_\gamma$  & = & $1.07$ & ${}^{+0.10}_{-0.10}$ & $\pm
    0.07$ &
    $0.76 \; <$ &  $\kappa_\gamma$ & $< \; 1.36 \;\; (95\% \,\mathrm{C.L.}) $
    \\[0.7ex]
    $\lambda_\gamma$ & = & $0.31$ & ${}^{+0.12}_{-0.20}$ & $\pm 0.07$ &
    $-0.45 \; <$ & $\lambda_\gamma$ & $< \; 0.70 \;\; (95\% \,\mathrm{C.L.}) $
    \,,
  \end{tabular}
\end{center}
with a correlation of $-12\%$. The corresponding 68\% and 95\%
confidence level contours are shown in Figure~\ref{fig:cn}.  These
results represent a considerable improvement in the accuracy compared
to our previous measurements~\cite{l3wen2} and are complementary to
those determined at the Tevatron~\cite{cdf-do} and from $\WW$
production at LEP~\cite{alephwen2,delphitgc,l3ww-opaltgc}, in
particular for the parameter $\kappa_\gamma$.

\section*{Appendix}
The results on the single W cross-section are also expressed in a
different phase space region to allow combination with other LEP
experiments. Single W production can alternatively be
defined as the complete $t$-channel subset of Feynman diagrams
contributing to the $\enu \FF$ final states with the following
kinematic cuts. For $\enu \QQ$ final states, the invariant mass of the
$\QQ$ pair is required to be greater than 45 \GeV. In the case of
$\enu\lan$, the energy of the lepton, $E_{\ell^-}$, must be greater
than 20 \GeV. In addition, for the $\enu\ean$ final state the
following angular cuts are applied: $|\cos{\theta_\mathrm{e^+}}|>0.95$
and $|\cos{\theta_\mathrm{e^-}}|<0.95$. The measured cross sections
corresponding to these phase space conditions are given in
Table~\ref{tab:lep}.

\clearpage

\typeout{   }     
\typeout{Using author list for paper 256 -  }
\typeout{$Modified: Jul 15 2001 by smele $}
\typeout{!!!!  This should only be used with document option a4p!!!!}
\typeout{   }
%
%
%
%
%
%

\newcount\tutecount  \tutecount=0
\def\tutenum#1{\global\advance\tutecount by 1 \xdef#1{\the\tutecount}}
\def\tute#1{$^{#1}$}
\tutenum\aachen            
\tutenum\nikhef            
\tutenum\mich              
\tutenum\lapp              
\tutenum\basel             
\tutenum\lsu               
\tutenum\beijing           
\tutenum\berlin            
\tutenum\bologna           
\tutenum\tata              
\tutenum\ne                
\tutenum\bucharest         
\tutenum\budapest          
\tutenum\mit               
\tutenum\panjab            
\tutenum\debrecen          
\tutenum\dublin            
\tutenum\florence          
\tutenum\cern              
\tutenum\wl                
\tutenum\geneva            
\tutenum\hefei             
\tutenum\lausanne          
\tutenum\lyon              
\tutenum\madrid            
\tutenum\florida           
\tutenum\milan             
\tutenum\moscow            
\tutenum\naples            
\tutenum\cyprus            
\tutenum\nymegen           
\tutenum\caltech           
\tutenum\perugia           
\tutenum\peters            
\tutenum\cmu               
\tutenum\potenza           
\tutenum\prince            
\tutenum\riverside         
\tutenum\rome              
\tutenum\salerno           
\tutenum\ucsd              
\tutenum\sofia             
\tutenum\korea             
\tutenum\purdue            
\tutenum\psinst            
\tutenum\zeuthen           
\tutenum\eth               
\tutenum\hamburg           
\tutenum\taiwan            
\tutenum\tsinghua          

{
\parskip=0pt
\noindent
{\bf The L3 Collaboration:}
\ifx\selectfont\undefined
 \baselineskip=10.8pt
 \baselineskip\baselinestretch\baselineskip
 \normalbaselineskip\baselineskip
 \ixpt
\else
 \fontsize{9}{10.8pt}\selectfont
\fi
\medskip
\tolerance=10000
\hbadness=5000
\raggedright
\hsize=162truemm\hoffset=0mm
\def\r{\rlap,}
\noindent

P.Achard\r\tute\geneva\ 
O.Adriani\r\tute{\florence}\ 
M.Aguilar-Benitez\r\tute\madrid\ 
J.Alcaraz\r\tute{\madrid,\cern}\ 
G.Alemanni\r\tute\lausanne\
J.Allaby\r\tute\cern\
A.Aloisio\r\tute\naples\ 
M.G.Alviggi\r\tute\naples\
H.Anderhub\r\tute\eth\ 
V.P.Andreev\r\tute{\lsu,\peters}\
% ...
\newpage
\begin{list}{A}{\itemsep=0pt plus 0pt minus 0pt\parsep=0pt plus 0pt minus 0pt
                \topsep=0pt plus 0pt minus 0pt}
\item[\aachen]
 I. Physikalisches Institut, RWTH, D-52056 Aachen, Germany$^{\S}$\\
 III. Physikalisches Institut, RWTH, D-52056 Aachen, Germany$^{\S}$
\item[\nikhef] National Institute for High Energy Physics, NIKHEF, 
     and University of Amsterdam, NL-1009 DB Amsterdam, The Netherlands
\item[\mich] University of Michigan, Ann Arbor, MI 48109, USA
\item[\lapp] Laboratoire d'Annecy-le-Vieux de Physique des Particules, 
     LAPP,IN2P3-CNRS, BP 110, F-74941 Annecy-le-Vieux CEDEX, France
\item[\basel] Institute of Physics, University of Basel, CH-4056 Basel,
     Switzerland
\item[\lsu] Louisiana State University, Baton Rouge, LA 70803, USA
\item[\beijing] Institute of High Energy Physics, IHEP, 
  100039 Beijing, China$^{\triangle}$ 
\item[\berlin] Humboldt University, D-10099 Berlin, Germany$^{\S}$
\item[\bologna] University of Bologna and INFN-Sezione di Bologna, 
     I-40126 Bologna, Italy
\item[\tata] Tata Institute of Fundamental Research, Mumbai (Bombay) 400 005, India
\item[\ne] Northeastern University, Boston, MA 02115, USA
\item[\bucharest] Institute of Atomic Physics and University of Bucharest,
     R-76900 Bucharest, Romania
\item[\budapest] Central Research Institute for Physics of the 
     Hungarian Academy of Sciences, H-1525 Budapest 114, Hungary$^{\ddag}$
\item[\mit] Massachusetts Institute of Technology, Cambridge, MA 02139, USA
\item[\panjab] Panjab University, Chandigarh 160 014, India.
\item[\debrecen] KLTE-ATOMKI, H-4010 Debrecen, Hungary$^\P$
\item[\dublin] Department of Experimental Physics,
  University College Dublin, Belfield, Dublin 4, Ireland
\item[\florence] INFN Sezione di Firenze and University of Florence, 
     I-50125 Florence, Italy
\item[\cern] European Laboratory for Particle Physics, CERN, 
     CH-1211 Geneva 23, Switzerland
\item[\wl] World Laboratory, FBLJA  Project, CH-1211 Geneva 23, Switzerland
\item[\geneva] University of Geneva, CH-1211 Geneva 4, Switzerland
\item[\hefei] Chinese University of Science and Technology, USTC,
      Hefei, Anhui 230 029, China$^{\triangle}$
\item[\lausanne] University of Lausanne, CH-1015 Lausanne, Switzerland
\item[\lyon] Institut de Physique Nucl\'eaire de Lyon, 
     IN2P3-CNRS,Universit\'e Claude Bernard, 
     F-69622 Villeurbanne, France
\item[\madrid] Centro de Investigaciones Energ{\'e}ticas, 
     Medioambientales y Tecnol\'ogicas, CIEMAT, E-28040 Madrid,
     Spain${\flat}$ 
\item[\florida] Florida Institute of Technology, Melbourne, FL 32901, USA
\item[\milan] INFN-Sezione di Milano, I-20133 Milan, Italy
\item[\moscow] Institute of Theoretical and Experimental Physics, ITEP, 
     Moscow, Russia
\item[\naples] INFN-Sezione di Napoli and University of Naples, 
     I-80125 Naples, Italy
\item[\cyprus] Department of Physics, University of Cyprus,
     Nicosia, Cyprus
\item[\nymegen] University of Nijmegen and NIKHEF, 
     NL-6525 ED Nijmegen, The Netherlands
\item[\caltech] California Institute of Technology, Pasadena, CA 91125, USA
\item[\perugia] INFN-Sezione di Perugia and Universit\`a Degli 
     Studi di Perugia, I-06100 Perugia, Italy   
\item[\peters] Nuclear Physics Institute, St. Petersburg, Russia
\item[\cmu] Carnegie Mellon University, Pittsburgh, PA 15213, USA
\item[\potenza] INFN-Sezione di Napoli and University of Potenza, 
     I-85100 Potenza, Italy
\item[\prince] Princeton University, Princeton, NJ 08544, USA
\item[\riverside] University of Californa, Riverside, CA 92521, USA
\item[\rome] INFN-Sezione di Roma and University of Rome, ``La Sapienza",
     I-00185 Rome, Italy
\item[\salerno] University and INFN, Salerno, I-84100 Salerno, Italy
\item[\ucsd] University of California, San Diego, CA 92093, USA
\item[\sofia] Bulgarian Academy of Sciences, Central Lab.~of 
     Mechatronics and Instrumentation, BU-1113 Sofia, Bulgaria
\item[\korea]  The Center for High Energy Physics, 
     Kyungpook National University, 702-701 Taegu, Republic of Korea
\item[\purdue] Purdue University, West Lafayette, IN 47907, USA
\item[\psinst] Paul Scherrer Institut, PSI, CH-5232 Villigen, Switzerland
\item[\zeuthen] DESY, D-15738 Zeuthen, Germany
\item[\eth] Eidgen\"ossische Technische Hochschule, ETH Z\"urich,
     CH-8093 Z\"urich, Switzerland
\item[\hamburg] University of Hamburg, D-22761 Hamburg, Germany
\item[\taiwan] National Central University, Chung-Li, Taiwan, China
\item[\tsinghua] Department of Physics, National Tsing Hua University,
      Taiwan, China
\item[\S]  Supported by the German Bundesministerium 
        f\"ur Bildung, Wissenschaft, Forschung und Technologie
\item[\ddag] Supported by the Hungarian OTKA fund under contract
numbers T019181, F023259 and T037350.
\item[\P] Also supported by the Hungarian OTKA fund under contract
  number T026178.
\item[$\flat$] Supported also by the Comisi\'on Interministerial de Ciencia y 
        Tecnolog{\'\i}a.
\item[$\sharp$] Also supported by CONICET and Universidad Nacional de La Plata,
        CC 67, 1900 La Plata, Argentina.
\item[$\triangle$] Supported by the National Natural Science
  Foundation of China.
\end{list}
}
\vfill


\clearpage
\begin{landscape}
\begin{table}[p]
 \begin{center}
 \renewcommand{\arraystretch}{1.1}
 \begin{tabular}{|l||c|c|c|c||c|c|c|c|}
 \hline
 & \multicolumn{4}{c ||}{$\sqrt{s} = 191.6\GeV$ \qquad ${\cal L} =
 29.7\,\pb$} &
 \multicolumn{4}{c |}{$\sqrt{s} = 195.5\GeV$ \qquad ${\cal L} = 83.7\,\pb$} \\
 \hline
 \hline
 Final State          &
 $N_{data}$ &
 $N_{MC}^{tot}$           &
 $N_{MC}^{sign}$ &
 $\varepsilon$ [\%] &
 $N_{data}$ &
 $N_{MC}^{tot}$           &
 $N_{MC}^{sign}$ &
 $\varepsilon$ [\%] \\
 \hline                                                                                
 \pz $\epl\nel\QQ$        &    26 & $   26.4 \pm 0.3 $ & $\pz 5.8 \pm 0.1 $   & $47.4$ 
                          &    92 & $   79.7 \pm 0.7 $ & $   17.6 \pm 0.2 $   & $48.6$ \\
 \hline
 \pz $\epl\nel\emi\ane$   & \pz 3 & $\pz 2.2 \pm 0.1 $ & $\pz 1.24 \pm 0.02 $ & $73.2$ 
                          & \pz 9 & $\pz 7.3 \pm 0.2 $ & $\pz 3.87 \pm 0.06 $ & $74.0$ \\
 \pz $\epl\nel\mu^-\anm$  & \pz 1 & $\pz 1.6 \pm 0.3 $ & $\pz 0.99 \pm 0.01 $ & $53.3$ 
                          & \pz 4 & $\pz 3.8 \pm 0.2 $ & $\pz 3.05 \pm 0.04 $ & $53.6$ \\
 \pz $\epl\nel\tau^-\ant$ & \pz 1 & $\pz 0.8 \pm 0.1 $ & $\pz 0.47 \pm 0.01 $ & $30.5$ 
                          & \pz 2 & $\pz 2.5 \pm 0.1 $ & $\pz 1.40 \pm 0.03 $ & $30.3$ \\
 \hline                                                                                
 \pz $\epl\nel\ell^-\anl$ & \pz 5 & $\pz 4.6 \pm 0.3 $ & $\pz 2.7 \pm 0.1 $   & $51.6$ 
                          &    15 & $   13.6 \pm 0.3 $ & $\pz 8.3 \pm 0.1 $   & $51.7$ \\
 \hline
 \multicolumn{9}{c}{ }\\
 \hline
 &
 \multicolumn{4}{c||}{$\sqrt{s} = 199.5\GeV$ \qquad ${\cal L} = 82.8\,\pb$} &
 \multicolumn{4}{c|}{$\sqrt{s} = 201.8\GeV$ \qquad ${\cal L} = 37.0\,\pb$} \\
 \hline
 \hline
 Final State          &
 $N_{data}$ &
 $N_{MC}^{tot}$           &
 $N_{MC}^{sign}$ &
 $\varepsilon$ [\%]                              &                              
 $N_{data}$ &
 $N_{MC}^{tot}$           &
 $N_{MC}^{sign}$ &
 $\varepsilon$ [\%]\\
 \hline                                                                                
 \pz $\epl\nel\QQ$        &    77 & $   82.4 \pm 0.8 $ & $   19.3 \pm 0.2 $   & $49.6$ 
                          &    46 & $   36.9 \pm 0.4 $ & $\pz 9.1 \pm 0.1 $   & $51.5$ \\
 \hline
 \pz $\epl\nel\emi\ane$   &    13 & $\pz 7.3 \pm 0.3 $ & $\pz 4.02 \pm 0.07 $ & $71.9$ 
                          & \pz 6 & $\pz 3.3 \pm 0.1 $ & $\pz 1.87 \pm 0.04 $ & $75.0$ \\
 \pz $\epl\nel\mu^-\anm$  & \pz 3 & $\pz 4.3 \pm 0.2 $ & $\pz 3.16 \pm 0.04 $ & $52.1$ 
                          & \pz 1 & $\pz 2.0 \pm 0.1 $ & $\pz 1.50 \pm 0.03 $ & $53.6$ \\
 \pz $\epl\nel\tau^-\ant$ & \pz 2 & $\pz 2.3 \pm 0.1 $ & $\pz 1.48 \pm 0.03 $ & $30.1$ 
                          & \pz 1 & $\pz 1.2 \pm 0.1 $ & $\pz 0.71 \pm 0.02 $ & $31.3$ \\
 \hline                                                                                
 \pz $\epl\nel\ell^-\anl$ &    18 & $   13.9 \pm 0.3 $ & $\pz 8.7 \pm 0.2 $   & $50.7$ 
                          & \pz 8 & $\pz 6.5 \pm 0.2 $ & $\pz 4.1 \pm 0.1 $   & $52.3$ \\
 \hline
 \multicolumn{9}{c}{ }\\
 \hline
 &
 \multicolumn{4}{c||}{$\sqrt{s} = 204.8\GeV$ \qquad ${\cal L} = 79.0\,\pb$} &
 \multicolumn{4}{c|}{$\sqrt{s} = 206.6\GeV$ \qquad ${\cal L} = 139.1\,\pb$} \\
 \hline
 \hline
 Final State          &
 $N_{data}$ &
 $N_{MC}^{tot}$           &
 $N_{MC}^{sign}$ &
 $\varepsilon$ [\%]                               &
 $N_{data}$                              &
 $N_{MC}^{tot}$           &
 $N_{MC}^{sign}$ &
 $\varepsilon$ [\%] \\
 \hline                                                                                 
 \pz $\epl\nel\QQ$        & \pz 79 & $\pz 88.4 \pm 1.0 $ & $   19.9 \pm 0.2 $ & $ 51.2$ 
                          &    163 & $   158.0 \pm 1.8 $ & $   38.1 \pm 0.4 $ & $ 52.9$ \\
 \hline
 \pz $\epl\nel\emi\ane$   & \pzz 7 & $\pzz 6.6 \pm 0.4 $ & $\pz 3.6 \pm 0.1 $ & $ 70.2$ 
                          & \pz 12 & $\pz 12.0 \pm 0.7 $ & $\pz 6.6 \pm 0.1 $ & $ 72.8$ \\
 \pz $\epl\nel\mu^-\anm$  & \pzz 2 & $\pzz 3.3 \pm 0.2 $ & $\pz 2.7 \pm 0.1 $ & $ 47.8$ 
                          & \pzz 9 & $\pzz 6.2 \pm 0.2 $ & $\pz 5.2 \pm 0.1 $ & $ 49.8$ \\
 \pz $\epl\nel\tau^-\ant$ & \pzz 4 & $\pzz 2.1 \pm 0.2 $ & $\pz 1.7 \pm 0.1 $ & $ 25.9$ 
                          & \pzz 4 & $\pzz 3.6 \pm 0.4 $ & $\pz 1.9 \pm 0.1 $ & $ 24.7$ \\
 \hline                                                                                 
 \pz $\epl\nel\ell^-\anl$ & \pz 13 & $\pz 12.0 \pm 0.5 $ & $\pz 8.0 \pm 0.1 $ & $ 46.7$   
                          & \pz 25 & $\pz 21.8 \pm 0.8 $ & $   13.7 \pm 0.2 $ & $ 47.8$ \\
 \hline
 \end{tabular}
 \caption{ \label{tab:list} 
   The number of selected candidates for single W boson production,
   $N_{data}$, compared to the total number of expected events,
   $N_{MC}^{tot}$, for each decay channel of the W boson. The expected
   number of signal events, $N_{MC}^{sign}$, and the selection
   efficiencies, $\varepsilon$, are also shown. The quoted
   uncertainties are due to Monte Carlo statistics.}
 \end{center}
\end{table}
\end{landscape}

\clearpage
\begin{landscape}
  \begin{table}[th]
    \begin{center}
      \renewcommand{\arraystretch}{1.1}
      \begin{tabular}{|l||c|c|c|c|c|c|}
        \hline
        \rule[-3mm]{0mm}{8mm} $\sqrt{s}$ &
        191.6~\GeV &
        195.5~\GeV &
        199.5~\GeV &
        201.8~\GeV &
        204.6~\GeV &
        206.6~\GeV \\
      \hline
      \hline
      \rule[-3mm]{0mm}{8mm}$\sigma_{\e \nu \QQ}$  & 
      $0.67^{+ 0.35}_{- 0.29}\pm 0.04$ & 
      $0.53^{+ 0.19}_{- 0.18}\pm 0.03$ & 
      $0.29^{+ 0.18}_{- 0.16}\pm 0.02$ & 
      $0.87^{+ 0.32}_{- 0.28}\pm 0.04$ & 
      $0.34^{+ 0.20}_{- 0.17}\pm 0.02$ & 
      $0.53^{+ 0.15}_{- 0.14}\pm 0.03$ \\
      \hline
      \rule[-3mm]{0mm}{8mm}$\sigma_{\e \nu \QQ}^\mathrm{GRC4F}$  &
      0.406 &
      0.435 &
      0.465 &
      0.480 &
      0.483 &
      0.496 \\ 
      \hline
      \rule[-3mm]{0mm}{8mm} $\sigma_{\e \nu \QQ}^\mathrm{EXCALIBUR}$  & 
      0.398 & 
      0.439 & 
      0.461 & 
      0.474 & 
      0.527 & 
      0.544 \\
        \hline
        \hline
        \rule[-3mm]{0mm}{8mm}  $\sigma_{\e \nu \ell \nu} $  & 
      $0.22^{+ 0.18}_{- 0.13}\pm 0.02$ & 
      $0.23^{+ 0.10}_{- 0.09}\pm 0.01$ & 
      $0.32^{+ 0.12}_{- 0.10}\pm 0.02$ & 
      $0.31^{+ 0.18}_{- 0.14}\pm 0.02$ & 
      $0.23^{+ 0.11}_{- 0.10}\pm 0.01$ & 
      $0.29^{+ 0.08}_{- 0.07}\pm 0.02$ \\
      \hline
      \rule[-3mm]{0mm}{8mm} $\sigma_{\e \nu \ell \nu}^\mathrm{GRC4F}$  &
      0.182 &
      0.196 &
      0.209 &
      0.215 &
      0.225 &
      0.231 \\
      \hline
      \rule[-3mm]{0mm}{8mm} $\sigma_{\e \nu \ell \nu}^\mathrm{EXCALIBUR}$  &
      0.195 & 
      0.213 & 
      0.229 & 
      0.232 & 
      0.237 & 
      0.243 \\
      \hline
      \hline
      \rule[-3mm]{0mm}{8mm} $\sigma_{\e \nu \W}$  & 
    $0.86^{+ 0.37}_{- 0.32}\pm 0.04$ & 
    $0.75^{+ 0.21}_{- 0.19}\pm 0.03$ & 
    $0.69^{+ 0.20}_{- 0.18}\pm 0.03$ & 
    $1.16^{+ 0.35}_{- 0.31}\pm 0.04$ & 
    $0.61^{+ 0.22}_{- 0.20}\pm 0.03$ & 
    $0.84^{+ 0.16}_{- 0.16}\pm 0.03$ \\
    \hline
    \rule[-3mm]{0mm}{8mm}  $\sigma_{\e \nu \W}^\mathrm{GRC4F}$  &
    0.588 &
    0.631 &
    0.674 &
    0.695 &
    0.721 &
    0.727 \\
    \hline
    \rule[-3mm]{0mm}{8mm}  $\sigma_{\e \nu \W}^\mathrm{EXCALIBUR}$  & 
    0.593 & 
    0.652 & 
    0.689 & 
    0.706 & 
    0.761 & 
    0.788 \\
    \hline
  \end{tabular}
  \caption{ \label{tab:xsection} 
    Measured cross sections in pb of the single W process at
    centre-of-mass energies between 192 \GeV\ and 207 \GeV. The
    results for hadronically and leptonically decaying W bosons, as
    well as their combination are shown. The first uncertainty is
    statistical and the second systematic. Also listed are the
    Standard Model predictions calculated with the GRC4F and 
    EXCALIBUR Monte Carlo programs. The theoretical predictions
    presented here are calculated with a statistical accuracy of
    $0.2\%-1.0\%$.  The current theoretical uncertainty on the single
    W cross section is of the order of 5\%~[25].  }
\end{center}
\end{table}
\end{landscape}

\begin{table}[th]
 \begin{center}
 \renewcommand{\arraystretch}{1.1}
 \begin{tabular}{|c||c|c|}
   \hline
   \rule[-3mm]{0mm}{8mm} Source of uncertainty & \multicolumn{2}{c|}{Final state} \\
   \cline{2-3}
   \rule[-3mm]{0mm}{8mm} & $\Wqq$ & $\Wln$ \\
   \hline
   \hline
   \rule[-3mm]{0mm}{8mm} Signal modelling                  & 3.2 & 2.1 \\
   \rule[-3mm]{0mm}{8mm} Lepton identification             & --- & 1.5 \\
   \rule[-3mm]{0mm}{8mm} Trigger efficiency                & --- & 2.3 \\
   \rule[-3mm]{0mm}{8mm} Neural network                    & 3.0 & --- \\
   \rule[-3mm]{0mm}{8mm} Signal Monte Carlo statistics     & 1.0 -- 1.2 & 1.6 --
   2.1 \\
   \rule[-3mm]{0mm}{8mm} Background Monte Carlo statistics & 1.1 -- 3.4 & 1.9 --
   6.0 \\
   \rule[-3mm]{0mm}{8mm} Background cross section          & 0.6 & 0.4 
   \\                                                                 
   \rule[-3mm]{0mm}{8mm} Variation of binning              & 1.0 & 1.5 
   \\
   
   \hline
   \hline
   \rule[-3mm]{0mm}{8mm} Total systematics                 & 4.8 -- 5.4 & 4.5 --
   7.3 \\
   \hline
 \end{tabular}
 \caption{ \label{tab:systematics} 
   Relative systematic uncertainties in per cent on the determination
   of the single W cross sections at $\sqrt{s}=192-209\GeV$ for the
   hadronic and leptonic final states. The uncertainties due to Monte
   Carlo statistics vary at the different centre-of-mass energies.  }
 \end{center}
\end{table}

\begin{table}[th]
 \begin{center}
 \renewcommand{\arraystretch}{1.1}
 \begin{tabular}{|c||c|c|c||c|c|c|}
 \hline
 \rule[-3mm]{0mm}{8mm} $\sqrt{s}$ &
 $\sigma_{\e \nu \QQ}$ &
 $\Delta\sigma_{stat}^{exp}$ & 
 $\sigma_{\e \nu \QQ}^\mathrm{GRC4F}$ &
 $\sigma_{\e\nu\W}$ & 
 $\Delta\sigma_{stat}^{exp}$ &
 $\sigma_{\e \nu \W}^\mathrm{GRC4F}$ 
 \\
 \hline
 \rule[-3mm]{0mm}{8mm}182.7~\GeV & $0.58^{+0.23}_{-0.20}\pm 0.04$ & $0.21$ &0.42 & $0.80^{+0.28}_{-0.25}\pm 0.05$ & $0.26$ & 0.63\\ \hline
 \rule[-3mm]{0mm}{8mm}188.6~\GeV & $0.52^{+0.14}_{-0.13}\pm 0.03$ & $0.14$ &0.46 & $0.69^{+0.16}_{-0.14}\pm 0.04$ & $0.15$ & 0.69\\ \hline
 \rule[-3mm]{0mm}{8mm}191.6~\GeV & $0.84^{+0.44}_{-0.37}\pm 0.04$ & $0.41$ &0.49 & $1.11^{+0.48}_{-0.41}\pm 0.05$ & $0.46$ & 0.73\\ \hline
 \rule[-3mm]{0mm}{8mm}195.5~\GeV & $0.66^{+0.24}_{-0.22}\pm 0.03$ & $0.21$ &0.52 & $0.97^{+0.27}_{-0.25}\pm 0.03$ & $0.25$ & 0.78\\ \hline
 \rule[-3mm]{0mm}{8mm}199.5~\GeV & $0.37^{+0.22}_{-0.20}\pm 0.02$ & $0.22$ &0.56 & $0.88^{+0.26}_{-0.24}\pm 0.04$ & $0.25$ & 0.84\\ \hline
 \rule[-3mm]{0mm}{8mm}201.8~\GeV & $1.10^{+0.40}_{-0.35}\pm 0.06$ & $0.35$ &0.58 & $1.50^{+0.45}_{-0.40}\pm 0.05$ & $0.38$ & 0.87\\ \hline
 \rule[-3mm]{0mm}{8mm}204.8~\GeV & $0.42^{+0.25}_{-0.21}\pm 0.03$ & $0.25$ &0.61 & $0.78^{+0.29}_{-0.25}\pm 0.04$ & $0.29$ & 0.91\\ \hline
 \rule[-3mm]{0mm}{8mm}206.6~\GeV & $0.66^{+0.19}_{-0.17}\pm 0.04$ & $0.20$ &0.62 & $1.08^{+0.21}_{-0.20}\pm 0.04$ & $0.23$ & 0.94\\ \hline
 \end{tabular}
 \caption{ \label{tab:lep} 
   Measured hadronic and total cross sections in pb at $\sqrt{s}=183-189
   \GeV$~[4,5] and at $\sqrt{s}=192-207 \GeV$ using an alternative
   signal definition of the single W process.  The first uncertainty
   is statistical and the second systematic. Also listed are the
   expected statistical uncertainties, $\Delta\sigma_{stat}^{exp}$, at
   each centre-of-mass energy and the Standard Model predictions
   calculated with GRC4F. }
 \end{center}
\end{table}

\clearpage
\vfill

\clearpage

\clearpage
\vfill

\clearpage
\vfill

\clearpage
\vfill

\clearpage
\vfill

\end{document}